\input{aipcheck}

\documentclass[
    ,final            % use final for the camera ready runs
  ]
  {aipproc}

\layoutstyle{6x9}

\begin{document}

\title{The Low Frequency Instrument in the ESA {\sc Planck} mission}

\author{A. Mennella}{
  address={IASF-CNR Sez. di Milano, Via Bassini 15, Milano (Italy)}
}

\author{M. Bersanelli}{
  address={Universit\'a degli Studi di Milano, Physics Department, Via Celoria 16, Milano (Italy)}
}

\author{B. Cappellini}{
  address={Universit\'a degli Studi di Milano, Physics Department, Via Celoria 16, Milano (Italy)}
}

\author{D. Maino}{
  address={Universit\'a degli Studi di Milano, Physics Department, Via Celoria 16, Milano (Italy)}
}

\author{P. Platania}{
  address={Universit\'a degli Studi di Milano, Physics Department, Via Celoria 16, Milano (Italy)}
}

\author{S. Garavaglia}{
  address={Universit\'a degli Studi di Milano, Physics Department, Via Celoria 16, Milano (Italy)}
}

\author{R.C. Butler}{
  address={IASF-CNR Sez. di Bologna, Via Gobetti 101, Bologna (Italy)}
}

\author{N. Mandolesi}{
  address={IASF-CNR Sez. di Bologna, Via Gobetti 101, Bologna (Italy)}
}

\author{F. Pasian}{
  address={INAF-OAT, Via Tiepolo 11, Trieste (Italy)}
}

\author{O. D'arcangelo}{
  address={IFP-CNR, Via Cozzi 53, Milano (Italy)}
}

\author{A. Simonetto}{
  address={IFP-CNR, Via Cozzi 53, Milano (Italy)}
}

\author{C. Sozzi}{
  address={IFP-CNR, Via Cozzi 53, Milano (Italy)}
}

\begin{abstract}

    Measurements of the cosmic microwave background (CMB) allow high precision observation of the 
    cosmic plasma at redshift $z\sim$1100. After the success of the NASA satellite COBE, that
    in 1992 provided the first detection of the CMB anisotropy, results from many ground-based and 
    balloon-borne experiments have showed a remarkable 
    consistency between different results and provided quantitative estimates of fundamental 
    cosmological properties \cite{Bersanelli:2002}. During the current year the team of the
    NASA WMAP satellite has released the first improved full-sky maps of the CMB since COBE, 
    leading to a deeper insight in the origin and evolution of the Universe.
    The ESA satellite {\sc Planck}, scheduled for launch in 2007, is designed to provide the ultimate 
    measurement of the CMB temperature anisotropy over the full sky, with an accuracy 
    that will be limited only by astrophysical foregrounds, and robust detection of polarisation
    anisotropy. {\sc Planck} will observe the sky with two instruments
    over a wide spectral band (the Low Frequency Instrument, based on coherent radiometers, 
    from 30 to 70 GHz and the 
    High Frequency Instrument, based on bolometric detectors, from 100 to 857 GHz). 
    The mission performances will improve dramatically the scientific return compared to
    WMAP.
    Furthermore the LFI radiometers (as well as some of the HFI 
    bolometers) are intrinsically sensitive to polarisation so that by combining the data from 
    different receivers it will be possible to measure 
    accurately the E mode and to 
    detect the B mode of the polarisation power spectrum. {\sc Planck} sensitivity will 
    offer also the possibility to detect the non-Gaussianities imprinted in the CMB.

\end{abstract}

\maketitle

%%%%%%%%%%%%%%%%%%%%%%%%%%%%%%%%%%%%%%%%%%%%
%% MAINMATTER
%%%%%%%%%%%%%%%%%%%%%%%%%%%%%%%%%%%%%%%%%%%%

%\section{Introduction}

\section{The {\sc Planck} satellite}

%    Satellite and Payload overview

\begin{figure}[here]
        \resizebox{13. cm}{!}{\includegraphics{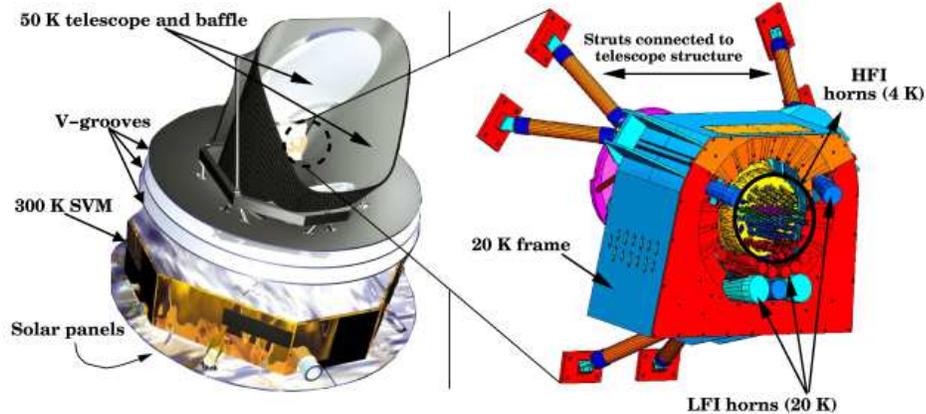}}
        \caption{                       
            Schematic picture of the {\sc Planck} satellite (left panel). Close-up of the instruments in
            the focal plane (right panel).
        }
        \label{fig:Planck_satellite}
    \end{figure}

    The {\sc Planck} satellite is shown in the left panel of 
    Fig.~\ref{fig:Planck_satellite}. Its design has been largely driven
    by its highly stringent thermal requirements, which include cold optics and cryogenically
    cooled instruments with highly stable thermal stages from 300~K down to 0.1~K. The satellite
    is composed by two main units:
    a warm ($\sim$300~K) Service Module (SVM) containing the warm electronics and the cooler
    compressor systems, and the Payload, passively cooled at $\sim$50~K with
    the off-axis dual reflector aplanatic telescope and the instruments in the focal plane. Three
    thermal shields at $\sim$150, 100 and 50~K (so-called ``V-grooves'') thermally decouple the Payload
    by the SVM by radiating heat into space.

    The two instruments in the focal plane (shown in the right panel of 
    Fig~\ref{fig:Planck_satellite}) are tightly integrated, with the HFI located in the centre
    part of the focal surface and the LFI antennas arranged in a circle around the HFI.

    Three cryogenic coolers cool the {\sc Planck} instruments: a 20~K Hydrogen Sorption
    Cooler cools the front-end of the LFI and provides a precooling stage for the HFI 4~K
    Stirling Cooler, which cools the HFI horns and precools the 0.1~K Dilution Cooler,  
    necessary to reach the working temperature of the bolometric detectors. The Sorption Cooler \cite{Bhandari:2000},
    in particular, is a vibration-less cooler that provides $\sim$ 1~W cooling power at 20~K by means of 
    six ``compressors'' containing a metal hydride that adsorbs and desorbs hydrogen depending on the absolute
    temperature, thus generating a constant high pressure hydrogen gas flow that expands and liquefies in a
    Joule Thomson valve.

%    Orbit and scanning strategy

    {\sc Planck} will orbit around the second lagrangian point (L2) of the Sun-Earth system, at about 1.5 MKm from Earth;
    this way the Sun and the Earth will always be aligned in the direction of the satellite
    spin axis with the telescope constantly in shade, thus greatly improving thermal stability. 
    The instruments will scan the sky at 1 rpm with the telescope
    axis pointing at $\sim$85$^\circ$ with respect to the spin axis; repointing will occur in discrete steps at the
    rate of $\sim$2.5' per hour. Following this strategy {\sc Planck} will cover the whole sky twice in about 14 months, 
    which corresponds to the current baseline mission time.

\section{The LFI}

    The LFI is a radiometer array in the 30-70~GHz range based on ultra-low noise
    indium phosphide (InP) high-electron-mobility transistors (HEMTs) cryogenically
    cooled at 20~K, which have shown world-record noise performances 
    over the required 20\% bandwidth with very low
    power dissipation (less that 0.5~W at 20~K) at LFI frequencies. 
    
\begin{figure}[here]
        \resizebox{14. cm}{!}{\includegraphics{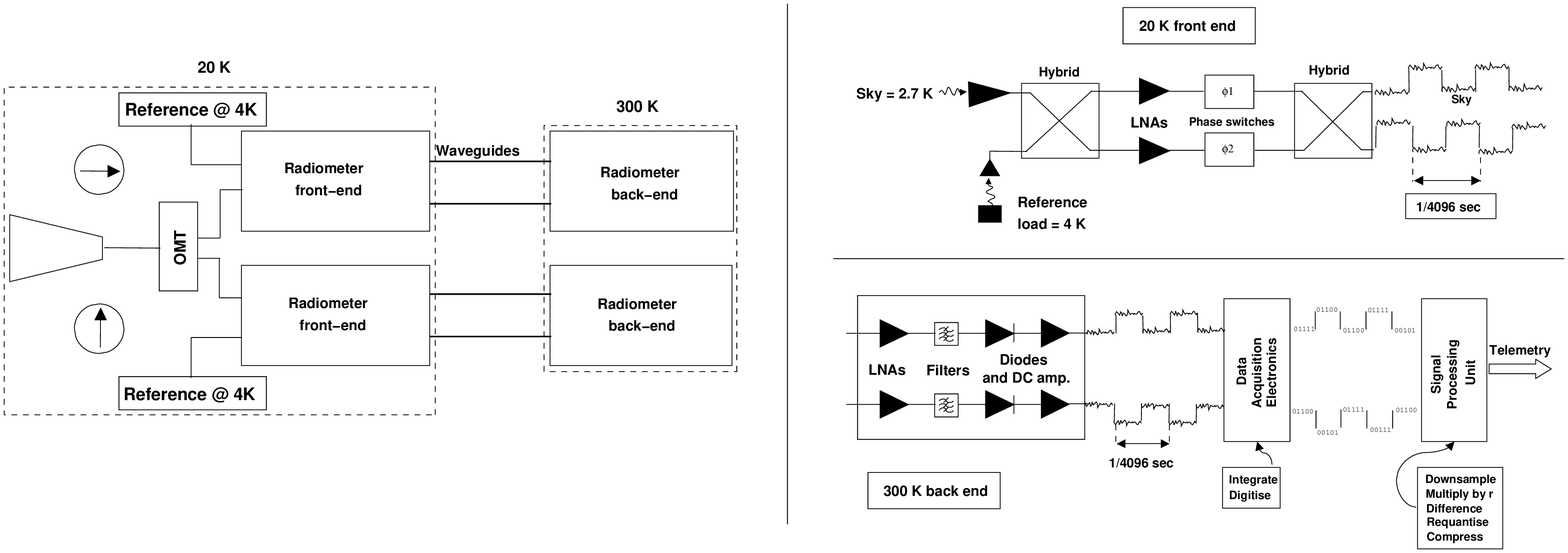}}
        \caption{                       
          Baseline LFI pseudo-correlation radiometer.
        }
        \label{fig:receiver}
    \end{figure}

    The particular pseudo-correlation differential
    design adopted for the LFI radiometers \cite{Seiffert:2002} (see Fig.~\ref{fig:receiver}) 
    ensures a very low susceptibility of the
    measured signal to 1/$f$ instabilities of the front-end amplifiers.
% which is essential
%    to guarantee a low systematic error contamination in the final measurements.
    
    In each receiver chain a corrugated dual profiled
    feed-horn collects and feeds the incoming radiation to an Orthomode Transducer
    (OMT) that separates the two perpendicularly polarised components. Each feed horn
    has been custom-designed to optimise the optical response in terms of angular resolution
    and sidelobe rejection. Every feed-horn in the {\sc Planck} focal plane 
    has a matched companion pointing in the same direction but with
    a relative orientation of 45$^\circ$ in the sky\footnote{Apart for one of the three 44~GHz
    feeds that could not be optimised because of mechanical constraints}. This disposition is
    optimal for the reconstruction of the CMB polarisation anisotropy.
    Further details about design and realisation of LFI feed-horns can be found in \cite{Villa:2003}.
    
    Each arm of the OMT output is connected to a radiometer composed by a 
    cooled front-end (20 K) and a 300 K back-end linked by waveguides. 
    In the baseline design the sky signal coming from each OMT arm is coupled to the reference 4~K signal 
    to the HEMT amplifiers via a 180$^\circ$ hybrid. One of the two signals then runs through a switch 
    that applies a phase shift which oscillates between 0 and 180$^\circ$ at a frequency 
    of 4096 Hz. A second phase switch is present for symmetry on the second radiometer 
    leg but does not introduce any phase shift in the propagating signal. The signals 
    are then recombined by a second 180$^\circ$ hybrid, producing an output which is a 
    sequence of signals alternating at twice the phase switch frequency. In the back-end 
    the RF signals are further amplified, filtered by a low-pass filter and then detected. 
    After detection the sky and reference load signals are integrated, digitised and then 
    differenced after multiplication of the reference load signal by a so-called  
    gain modulation factor , $r$, which has the function to make the sky-load 
    difference as close as possible to zero and cancel at first order 1/$f$ instabilities
    from the RF radiometer components.

    In Tab.~\ref{tab:LFI_performances} we report a summary of the LFI main scientific
    performances. 

    \begin{table}
    \begin{tabular}{lrrr}
    \hline
    LFI frequency (GHz)                                 & 30     & 44        & 70       \\
    Noise per 30' pixel ($\mu$K)                        & 6      & 6         & 6        \\
    1-sec sensitivity (mK$\times$s$^{1/2}$)             & 179    &    219    &   310    \\
    Number of feeds                                     &  2     &   3       &  6       \\
    System Temperature (K)                              &  7.5   &      12   &  21.5    \\
    Effective bandwidth                                 &  20\%  &      20\% &  20\%    \\
    Angular resolution                                  &  30'   &      24'  &  14'     \\
    Av. $\Delta T/T$ per resolution element (10$^{-6}$)                                 \\
    -total intensity-                                   &  2.2   &   2.8     &  4.9     \\
    -polarisation-                                      &  3.1   &   3.9     &  6.9     \\
    \hline
    \end{tabular}
    \caption{Goal performances of LFI radiometers}
    \label{tab:LFI_performances}
    \end{table}

\section{Systematic error control}

    The extraordinary sensitivity of {\sc Planck} instruments calls for consistently tight requirements
    on systematic errors, especially those  that are synchronous 
    with the spacecraft spin period. Spin-synchronous signals, in fact, are virtually 
    indistinguishable from any true sky signal and, therefore, will affect permanently 
    the final LFI maps. 

    In general we can classify systematic errors into five broad categories: 

    \begin{description}
        \item[main Beam distortions,] i.e. deviations 
        of the real main beams from perfect Gaussian beams
        and differences between beam shapes at the same frequency;

        \item[instrument Intrinsic effects,] i.e. systematic effects that arise from non-idealities 
        of the radiometers themselves (e.g. amplifier 1/$f$ noise); 

        \item[external Straylight,] i.e. signals coming 
        from celestial sources entering the instrument through the antenna side-lobes;

        \item[pointing errors,] which depend on the limitations of the satellite attitude control subsystem 
        and on the ability to reconstruct the orientation of the beam using star sensor and instrument data;

        \item[thermal Effects,] caused by thermal oscillations on time-scales 
        ranging from less than a minute to more than an hour. 

    \end{description}

    Both the satellite and the instruments 
    are designed to maintain the contamination from astrophysical
    and instrumental systematic error at the level of few $\mu$K per pixel in the final maps. This low level
    of contamination will be guaranteed by the very stable thermal environment in L2, a careful
    thermal engineering of the thermal interfaces with the intruments and by advanced data analysis 
    procedures that will fully exploit the measurement redundancies provided by the scanning strategy 
    \cite{Mennella:2002, Maino:2002}. Such data reduction steps are
    planned to be carried out as part of the LFI Data Processing Centre
    (DPC) activities \cite{Pasian:2002}.

%%%%%%%%%%%%%%%%%%%%%%%%%%%%%%%%%%%%%%%%%%%%
%% SAMPLE TABLE
%%
%% Shows the use of \tablehead and \tablenote
%% macros
%%%%%%%%%%%%%%%%%%%%%%%%%%%%%%%%%%%%%%%%%%%%

%\begin{description}
%\item[Infandum]
% regina, iubes renovare dolorem, Troianas ut opes et lamentabile
% regnum cruerint Danai.
%\item[Sed]
% si tantus amor casus cognoscere nostros et breviter Troiae supremum
% audire laborem, quamquam animus meminisse horret, luctuque refugit,
% incipiam.
%\item[Lamentabile] regnum cruerint Danai; quaeque ipse miserrima vidi, et
%quorum pars magna fui. Quis talia  fando Myrmidonum Dolopumve aut duri
%miles Ulixi temperet a lacrimis?
%\end{description}

%%%%%%%%%%%%%%%%%%%%%%%%%%%%%%%%%%%%%%%%%%%%%%%%
%% BACKMATTER
%%%%%%%%%%%%%%%%%%%%%%%%%%%%%%%%%%%%%%%%%%%%%%%%

%\begin{theacknowledgments}
%  Infandum, regina, iubes renovare dolorem, Troianas ut opes et
%  lamentabile regnum cruerint Danai; quaeque ipse miserrima vidi, et
%  quorum pars magna fui. Quis talia fando Myrmidonum Dolopumve aut duri
%  miles Ulixi temperet a lacrimis?
%\end{theacknowledgments}

%% If the bibliography is produced without BibTeX comment out the

\bibliographystyle{aipproc}   % if natbib is available

\bibliography{LFI_in_Planck_mission}

%\begin{thebibliography}{bibliography}

\IfFileExists{\jobname.bbl}{}
 {\typeout{}
  \typeout{******************************************}
  \typeout{** Please run "bibtex \jobname" to optain}
  \typeout{** the bibliography and then re-run LaTeX}
  \typeout{** twice to fix the references!}
  \typeout{******************************************}
  \typeout{}
 }

\end{document}